\RequirePackage{fix-cm}
\documentclass[smallextended]{svjour3wide}  
\smartqed  
\usepackage{graphicx}
\usepackage{amsmath}
\usepackage{amssymb}
\usepackage{color}
\usepackage{bm}
\journalname{Journal of Statistical Physics}
\begin{document}

\title{Nonequilibrium Langevin equation and effective temperature for particle interacting with spatially extended environment}
\titlerunning{Nonequilibrium Langevin equation for a driven Brownian particle} 

\author{Taiki Haga}

\institute{ \at
              Department of Physics, Kyoto University, Kyoto 606-8502, Japan \\
              \email{haga@scphys.kyoto-u.ac.jp}           
}

\date{Received: date / Accepted: date}

\maketitle

\begin{abstract}
We investigate a novel type of Langevin model that describes the nonequilibrium dynamics of a classical particle interacting with a spatially extended environment. 
In this model, a particle, which interacts with the environment through the nonlinear interaction Hamiltonian, is driven by a constant external force, and subsequently,
it reaches a nontrivial nonequilibrium steady state. 
We derive an effective Langevin equation for the particle in the nonequilibrium steady states.
Using this equation, we calculate the effective temperature defined as the ratio of the correlation function of the velocity fluctuation to 
the linear response function with respect to a small perturbation.
As a result, it is shown that the effective temperature associated with the time scale of the particle is identical to the kinetic temperature 
if the time scale of the environment and that of the particle are well separated.
Furthermore, a noteworthy expression, which relates the kinetic temperature with the curvature of the driving force-mean velocity curve, is derived.
\end{abstract}

\section{Introduction}

It is of interest both from the fundamental aspect and a practical viewpoint to understand the nonequilibrium dynamics of a particle interacting with a spatially extended environment.
The fact that this subject is deeply related to such diverse areas as condensed matter physics [1], chemical physics [2] and biological physics [3] has attracted the attention of
theoretical physicists. 
Recent progress in experimental technique enables us to access the detailed dynamics of colloidal particles immersed in a passive or active medium in the presence of an external driving force [4-6].
In spite of significant theoretical and experimental efforts towards understanding these dynamics, many aspects of this problem are still unclear.

For equilibrium cases, the Langevin equation has been a powerful tool for investigating the stochastic dynamics of a particle in a homogeneous environment.
The effect of the environment can be described by introducing a linear friction with memory and Gaussian colored noise
\begin{equation}
\frac{{\rm d}p(t)}{{\rm d}t}=-\int_{-\infty}^t \Gamma(t-s)\dot{q}(s){\rm d}s+R(t),
\end{equation}
where $q(t)$ and $p(t)$ denote the position and momentum of the particle, respectively. 
Here, the zero-mean Gaussian noise $R(t)$ satisfies the fluctuation-dissipation relation of the second kind
\begin{equation}
\left<R(t)R(t') \right>=k_{\rm B} T\Gamma(|t-t'|).
\end{equation}
The friction kernel $\Gamma(t)$ reflects the dynamical properties of the environment.
The validity of the Langevin equation has been confirmed experimentally and theoretically for some thermal or athermal systems [7-9].

It is not clear whether the Langevin description is applicable for the dynamics of a particle driven by an external force.
In such cases, since the environment may not be in equilibrium, the friction kernel $\Gamma(t)$ and the random force $R(t)$ are modified from those in equilibrium.
In particular there is no guarantee for the validity of the fluctuation-dissipation relation of the second kind Eq.~(2) in this situation.
For example, in Refs.~[10] and [11] the relation between the friction and the correlation of the random force is discussed for a Brownian particle suspended in a nonequilibrium medium.

The effective temperature is a useful ingredient to describe the violation of the fluctuation-dissipation relation for nonequilibrium states.
It is defined as the ratio of the correlation function of fluctuation to the linear response function with respect to small perturbations.
It is known that driven systems with two well-separated time scales, such as sheared glassy liquids, have two effective temperatures associated with each time scale [12-15].
In this case, the effective temperature associated with the slow time scale is identical to the kinetic temperature of a coupled Hamiltonian subsystem whose reaction time is equal to the time scale [12].
A similar result is obtained for a heated nanoparticle suspended in a fluid by using a detailed molecular dynamics simulation in Ref.~[16].
It is natural to expect that this relationship between the time-scale-dependent effective temperature and the kinetic temperature holds for a particle driven by an external force.

The main purpose of this research is to investigate the relationship between the effective temperature and the kinetic temperature for a particle interacting with an environment 
by introducing a microscopic model in which the dynamics of the environment are explicitly considered.
The system under study consists of a particle and an elastic medium that corresponds to a spatially extended environment.
The elastic medium couples with an external heat bath of temperature $T$.
Keeping a classical particle interacting with the lattice oscillation of a crystalline solid in mind, we propose the simplest model for describing the motion of the particle.
For this model, we derive an effective Langevin equation by eliminating the degrees of freedom of the environment.
The nonlinear interaction leads to a friction kernel $\Gamma(t)$ and a non-Gaussian random force $R(t)$ that explicitly depend on an external force in contrast to the ordinary Langevin formulation.
Using the effective Langevin equation, we calculate the kinetic temperature $T_{\rm K}$ in nonequilibrium steady states (NESSs) driven by a constant external force. 
The time scale dependent effective temperature $T_{\rm eff}(\omega)$ is also calculated from the correlation function of the velocity fluctuation and the linear response function.
As a result, it is shown that the effective temperature associated with the time scale of the particle is identical to the kinetic temperature 
if the time scale of the environment $\tau_{\rm E}$ and that of the particle $\tau_{\rm P}$ are well separated. 
Remarkably, we found that the kinetic temperature $T_{\rm K}$ is not equal to the external heat bath temperature $T$ 
even in the limit $\tau_{\rm P}/\tau_{\rm E} \to \infty $, where the environment is expected to behave as an ideal heat bath. 
Furthermore, we derive a simple formula Eq.~(55) that relates the kinetic temperature with the curvature of the driving force-mean velocity curve.

This paper is organized as follows. In Sec.~2, we introduce the model considered in this research. 
In Sec.~3, the effective Langevin equation for the particle is derived by eliminating the degrees of freedom of the environment. 
The correlation function of velocity fluctuation and the kinetic temperature are calculated by using the equation. 
In Sec.~4, we discuss the relationship between the effective temperature and the kinetic temperature in nonequilibrium steady states. 
We also derive a simple formula for the kinetic temperature.
In Sec.~5, we summarize this work.

\section{Model}

We consider a one-dimensional system consisting of a particle and an elastic medium that corresponds to a spatially extended environment. 
The position of the particle and the displacement field of the elastic medium are denoted by $q(t)$ and $u(x,t)$, respectively. 
For simplicity, we consider only longitudinal displacement of the elastic medium.
The total Hamiltonian is written as
\begin{equation}
H=H_{\rm P}+H_{\rm E}+H_{{\rm int}},
\end{equation}
where $H_{\rm P}$, $H_{\rm E}$, and $H_{\rm int}$ represent the Hamiltonians for the particle, elastic medium, and interaction between them, respectively. 
The dynamics of this system with an external force $F$ are described by the equation of motion
\begin{equation}
\frac{{\rm d}p}{{\rm d}t}=-\frac{\partial H}{\partial q}+F,
\end{equation}
for the particle and
\begin{equation}
\frac{\partial^2 u(x)}{\partial t^2}=-\frac{\delta H}{\delta u(x)}-\gamma \dot{u}(x)+\xi(x,t),
\end{equation}
for the elastic medium. Here, $p(t)=M \dot{q} (t)$ is momentum of the particle, $\gamma$ is a damping constant, and $\xi(x,t)$ is Gaussian zero-mean white noise satisfying
\begin{equation}
\left<\xi(x,t)\xi(x',t')\right>=2\gamma k_{\mathrm B} T\delta(x-x')\delta(t-t').
\end{equation}
The second and third terms of the right-hand side of Eq.~(5) express the effects of an external heat bath of temperature $T$.

The Hamiltonian of the particle is given by
\begin{equation}
H_{\rm P}=\frac{p^2}{2M}.
\end{equation}
For the Hamiltonian of the elastic medium, we assume the following simplest form:
\begin{equation}
H_{\rm E}=\frac{1}{2} \int \left\{\dot{u}(x)^2+Ku(x)^2+g\left( \frac{\partial  u(x)}{\partial x} \right)^2  \right\} {\rm d}x.
\end{equation}
This form is rewritten as
\begin{equation}
H_{\rm E}=\frac{1}{2} \int \left\{|\dot{u}_k|^2+\omega_k^2 |u_k|^2 \right\}\frac{{\rm d}k}{2\pi},
\end{equation}
in terms of Fourier components
\begin{eqnarray}
u_k=\int u(x) e^{-ikx} {\rm d}x, \nonumber
\end{eqnarray}
where $\omega_k=\sqrt{c_{\rm s}^2k^2+K}$, and $c_{\rm s}=\sqrt{g}$ denotes the sound velocity of the elastic medium. 
Thus, the environment is an assembly of noninteracting harmonic oscillators.

Let us select the form of the interaction Hamiltonian such that it satisfies the following conditions: 
\begin{enumerate}
\item $H_{\rm int}$ must have translational symmetry,
\begin{eqnarray}
H_{{\rm int}}(q+a,\{u(x)\})=H_{{\rm int}}(q,\{u(x+a)\}). \nonumber
\end{eqnarray}
\item $H_{\rm int}$ must have inversion symmetry,
\begin{eqnarray}
H_{{\rm int}}(q,\{u(x)\})=H_{{\rm int}}(-q,\{-u(-x)\}). \nonumber
\end{eqnarray}
\item $H_{{\rm int}}$ should be linear with respect to $u(x)$.
\end{enumerate}
From the above conditions, we obtain
\begin{equation}
H_{{\rm int}}=\lambda \int V_{{\rm int}}(q-x)({\rm L}u)(x){\rm d}x,
\end{equation}
where $V_{{\rm int}}(x)$ denotes an even function that rapidly decreases over the length scale $a$ and ${\rm L}$ denotes a linear operator consisting of odd order spatial derivatives. 
It is reasonable to adopt the simplest form
\begin{equation}
H_{{\rm int}}=-\lambda \int V_{{\rm int}}(q-x)\frac{\partial u(x)}{\partial x} {\rm d}x.
\end{equation}
The physical interpretation of this expression is simple. 
Since $u(x)$ can be regarded as the displacement of particles of the elastic medium whose equilibrium position is $x$, 
$-\partial u(x)/\partial x$ corresponds to the density fluctuation that should couple with the potential energy. 
This Hamiltonian is known as the Fr\"{o}hlich type interaction that describes the electron-phonon interaction in metals and it has been studied for a long time in context of the polaron problem [17].

The equations of motion are given by
\begin{equation}
\frac{{\rm d}p}{{\rm d}t}=\lambda \int \frac{\partial}{\partial q} V_{\rm int}(q-x) \frac{\partial u(x)}{\partial x} {\rm d}x+F,
\end{equation}
for the particle and
\begin{equation}
\frac{\partial^2 u(x)}{\partial t^2}=\left(g\frac{\partial^2}{\partial x^2}-K \right) u(x)  +\lambda \frac{\partial}{\partial q} V_{\rm int}(q-x)-\gamma \dot{u}(x)+\xi(x,t),
\end{equation}
for the elastic medium.

\section{Langevin equation for NESS}

Let us derive the effective equation of motion for the particle by eliminating the degrees of freedom of the environment. 
Eqs.~(12) and (13) are respectively rewritten in terms of the Fourier components as
\begin{equation}
\frac{{\rm d}p}{{\rm d}t}=-\lambda \int k^2 \tilde{V}_{\rm int}(k) u_k e^{ikq} \frac{{\rm d}k}{2\pi} +F,
\end{equation}
and
\begin{equation}
\frac{{\rm d}^2 u_k}{{\rm d}t^2}=-\omega_k^2 u_k-\gamma \dot{u}_k-ik\lambda \tilde{V}_{\rm int}(k)e^{-ikq}+\xi_k(t),
\end{equation}
where $\omega_k=\sqrt{c_{\mathrm s}^2k^2+K}$ and $\tilde{V}_{\rm int}(k)$ denotes the Fourier transform of the interaction potential $V_{\rm int}(x)$.
The Gaussian noise $\xi_k(t)$ satisfies
\begin{equation}
\left<\xi_k(t)\xi^*_{k'}(t') \right>=4\pi\gamma k_{\rm B} T\delta(k-k')\delta(t-t').
\end{equation}
The solution of Eq.~(15) can be obtained as
\begin{equation}
u_k(t)=-\frac{1}{\sqrt{\gamma^2-4\omega_k^2}}\int^t_{-\infty} {\rm d}s  \left(-e^{\alpha_{k+}(t-s)}+e^{\alpha_{k-}(t-s)} \right) \left(-ik \lambda  \tilde{V}_{\rm int}(k) e^{-ikq(s)}+\xi_k(s) \right),
\end{equation}
where $\alpha_{k\pm}=\frac{1}{2}(-\gamma \pm \sqrt{\gamma^2-4\omega_k^2})$.
Since we are interested in steady states, the terms that depend on the initial condition can be dropped.
Substituting Eq.~(17) into Eq.~(14) and integrating by parts, we obtain the following equation of motion for the particle:
\begin{equation}
\frac{{\mathrm d}p(t)}{{\mathrm d}t}=-\int_{-\infty}^t \Gamma(t,s)\dot{q}(s){\mathrm d}s+F+R(t),
\end{equation}
where the friction kernel $\Gamma(t,s)$ is defined as
\begin{equation}
\Gamma(t,s)=\lambda^2 \int \frac{{\mathrm d}k}{2\pi} \frac{k^4}{\omega_k^2 \sqrt{\gamma^2-4\omega_k^2}} \tilde{V}_{\rm int}(k)^2 (\alpha_{k+}e^{\alpha_{k-}(t-s)}-\alpha_{k-}e^{\alpha_{k+}(t-s)})e^{ik(q(t)-q(s))}.
\end{equation}
The random force $R(t)$ is given by
\begin{equation}
R(t)=-\lambda \int \frac{{\mathrm d}k}{2\pi} k^2 \tilde{V}_{\rm int}(k) e^{ikq(t)} u_k^{\lambda=0}(t),
\end{equation}
where $u_k^{\lambda=0}$ is $u_k$ without the interaction
\begin{equation}
u_k^{\lambda=0}(t)=-\frac{1}{\sqrt{\gamma^2-4\omega_k^2}} \int_{-\infty}^t (-e^{\alpha_{k+}(t-s)}+e^{\alpha_{k-}(t-s)})\xi_k(s){\rm d}s.
\end{equation}
Although the expression Eq.~(18) resembles the Langevin equation given by Eq.~(1), it should be noted that $\Gamma(t,s)$ and $R(t)$ are nonlinear functions of the particle's state $q(t)$. 
Therefore, the dynamics of the particle are described by the nonlinear Langevin equation with multiplicative non-Gaussian noise.
This means that this model exhibits nontrivial nonequilibrium dynamics in the presence of a non-zero driving force $F$.

We estimate the friction kernel $\Gamma(t,s)$ for NESS driven by the constant external force $F$.
The equation is rewritten as
\begin{equation}
M \frac{{\mathrm d}v(t)}{{\mathrm d}t}=-\int_{-\infty}^t {\mathrm d}s \int \frac{{\rm d}k}{2\pi} G_k(t-s)\cos k(q(t)-q(s))v(s)+F+R(t),
\end{equation}
where $G_k(t)$ is defined as
\begin{equation}
G_k(t)=\lambda^2 \frac{k^4}{\omega_k^2 \sqrt{\gamma^2-4\omega_k^2}} \tilde{V}_{\rm int}(k)^2 (\alpha_{k+}e^{\alpha_{k-}t}-\alpha_{k-}e^{\alpha_{k+}t}).
\end{equation}
We write the particle velocity as $v(t)=v_{\rm s}+\delta v(t)$, where $v_{\rm s}$ denotes the mean velocity, and we linearize Eq.~(22) with respect to the fluctuation $\delta v$.
Let us consider the condition that this approximation is valid.
We notice that the relaxation time of $G_k(t)$ is about $\gamma^{-1}$.
Furthermore, the dominant contribution of the $k$ integral in Eq.~(22) comes from a region near $k=1/a$.
Thus, we have the following condition:
\begin{equation}
\sqrt{\left< \delta v^2 \right>} \ll a \gamma,
\end{equation}
where $a$ denotes the interaction range of $V_{\rm int}(x)$.
We define $\Gamma^{(0)}(t)$ as
\begin{equation}
\Gamma^{(0)}(t)=\int \frac{{\rm d}k}{2\pi} G_k(t)\cos kv_{\rm s}t,
\end{equation}
which is obtained by simply replacing $k(q(t)-q(s))$ with its mean value $kv_{\rm s}(t-s)$ in $\Gamma(t,s)$.
Taking the average of Eq.~(22) for NESS, we have
\begin{equation}
F=v_{\rm s} \int^{\infty}_0 \Gamma^{(0)}(t) {\rm d}t+O(\delta v^2).
\end{equation}
Since $\int^{\infty}_0 \Gamma^{(0)}(t) {\rm d}t$ corresponds to the friction constant in NESS, we call it static friction.
Substituting $q(t)-q(s)=v_{\rm s}(t-s)+\int^t_s \delta v(\tau) {\rm d}\tau$ into Eq.~(22) and neglecting the quadratic term of the velocity fluctuation, we have
\begin{equation}
M \frac{{\mathrm d}}{{\mathrm d}t}\delta v(t)=-\int_{-\infty}^t {\mathrm d}s  \Gamma^{(0)}(t-s) \delta v(s)-\int_{-\infty}^t {\rm d}s \Gamma^{(1)}(t-s) \int^t_s \delta v(\tau) {\rm d}\tau +R(t),
\end{equation}
where $\Gamma^{(1)}(t)$ is defined as
\begin{equation}
\Gamma^{(1)}(t)=-\int \frac{{\rm d}k}{2\pi} G_k(t)kv_{\rm s}\sin kv_{\rm s}t.
\end{equation}
The second term of Eq.~(27), which vanishes for $F=0$, represents the correction due to nonlinear friction.
Introducing the Fourier transform
\begin{equation}
\delta \tilde{v}(\omega)=\int^{\infty}_{-\infty} \delta v(t) e^{-i \omega t} {\rm d}t,\:\:\:\:\:\: \tilde{\Gamma}^{(i)}(\omega)=\int^{\infty}_0 \Gamma^{(i)}(t)e^{-i \omega t} {\rm d}t,  \nonumber
\end{equation}
Eq.~(27) is rewritten as 
\begin{equation}
iM \omega \delta \tilde{v}(\omega)=-\tilde{\Gamma}_{\rm eff}(\omega) \delta \tilde{v}(\omega)+\tilde{R}(\omega),
\end{equation}
where $\tilde{\Gamma}_{\rm eff}(\omega)$ is defined as
\begin{equation}
\tilde{\Gamma}_{\rm eff}(\omega)=\tilde{\Gamma}^{(0)}(\omega)+\frac{1}{i \omega} \left[ \tilde{\Gamma}^{(1)}(0)-\tilde{\Gamma}^{(1)}(\omega) \right].
\end{equation}
This is the effective Langevin equation for the particle in NESS.
The effective friction kernel $\Gamma_{\rm eff}(t)$ is obtained by the inverse Fourier transform
\begin{equation}
\Gamma_{\rm eff}(t)=\int {\rm Re} \tilde{\Gamma}_{\rm eff}(\omega) e^{i \omega t} \frac{d \omega}{\pi}.
\end{equation}

We are interested in the following power spectrum of the velocity fluctuation $\delta v(t)=v(t)-v_{\mathrm s} $ in NESS:
\begin{equation}
C_{\rm ss}(\omega)=\int \left<\delta v(t)\delta v(0) \right>e^{-i\omega t}{\mathrm d}t.
\end{equation}
In order to calculate the power spectrum, it is required to know the correlation function of the random force $R(t)$.
We have from Eq.~(20)
\begin{equation}
\left< R(t)R^*(t') \right>=\lambda^2 \int \frac{{\rm d}k {\rm d}k'}{4 \pi^2} k^2 k'^2 \tilde{V}_{\rm int}(k)\tilde{V}_{\rm int}(k')\left< u_k^{\lambda=0}(t) u_{-k'}^{\lambda=0}(t')e^{ik(q(t)-q(t'))} \right>.
\end{equation}
We expand $e^{ik(q(t)-q(s))}$ with respect to $\delta v$, and consequently, it can be shown that
\begin{equation}
\left< R(t)R^*(t') \right>=\lambda^2 \int \frac{{\rm d}k {\rm d}k'}{4 \pi^2} k^2 k'^2 \tilde{V}_{\rm int}(k)\tilde{V}_{\rm int}(k')\left< u_k^{\lambda=0}(t) u_{-k'}^{\lambda=0}(t') \right>e^{ikv_{\rm s}(t-t')}+Rem.
\end{equation}
In order to estimate the correction term $Rem$ in Eq.~(34), we note that the relaxation time of the correlation function of $ u_k^{\lambda=0}(t)$,
\begin{equation}
\left< u_k^{\lambda=0}(t) u_{-k'}^{\lambda=0}(0) \right>=\frac{2 \pi k_{\rm B}T}{\omega_k^2} \frac{1}{\sqrt{\gamma^2-4\omega_k^2}}\left(\alpha_{k+}e^{\alpha_{k-}t}-\alpha_{k-}e^{\alpha_{k+}t}  \right)\delta(k-k'),
\end{equation}
is about $\gamma^{-1}$.
Furthermore, the dominant contribution of the $k$ integral in Eq.~(33) comes from a region near $k=1/a$.
Thus, we have $Rem=O\left(\left< \delta v^2 \right>/a^2 \gamma^2 \right)$.
Since the correction term can be dropped because of the condition (24), we obtain the correlation of the random force as
\begin{equation}
\left<R(t)R^*(t') \right>=k_{\mathrm B} T\Gamma^{(0)}(|t-t'|).
\end{equation}
From this result, we also obtain the following expression for $C_{\rm ss}(\omega)$:
\begin{equation}
C_{\rm ss}(\omega)=\frac{2k_{\mathrm B}T\:{\rm Re}\tilde{\Gamma}^{(0)}(\omega)}{{\rm Re}\tilde{\Gamma}_{\rm eff}(\omega)^2+(M\omega+{\rm Im}\tilde{\Gamma}_{\rm eff}(\omega))^2}.
\end{equation}
Furthermore, the kinetic temperature defined by
\begin{eqnarray}
k_{\mathrm B}T_{\mathrm K}=M \left <\delta v(t)^2 \right>
\end{eqnarray}
is given by
\begin{equation}
k_{\mathrm B}T_{\mathrm K}=\int \frac{{\rm d}\omega}{2\pi} \frac{2 M k_{\mathrm B}T\:{\rm Re}\tilde{\Gamma}^{(0)}(\omega)}{{\rm Re}\tilde{\Gamma}_{\rm eff}(\omega)^2+(M\omega+{\rm Im}\tilde{\Gamma}_{\rm eff}(\omega))^2}.
\end{equation}
Note that $T_{\rm K}$ is not equal to the bath temperature $T$.

\begin{figure}
 \centering
 \includegraphics[width=80mm]{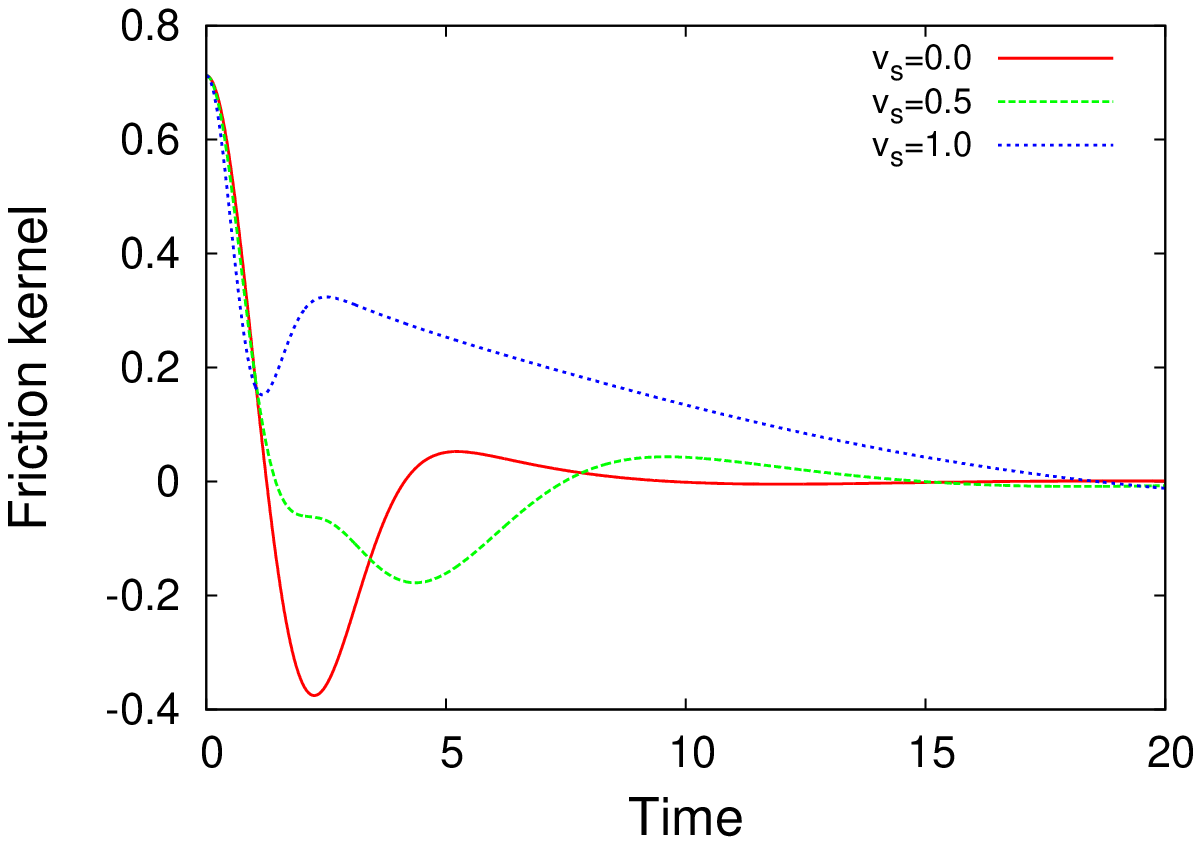}
 \caption{
Friction kernel $\Gamma_{\rm eff} (t)$ for the weak-friction limit for the mean velocity $v_{\rm s}$ = 0.0 (solid line), 0.5 (dashed line), and 1.0 (dotted line).
 Parameters: $\lambda=1$, $c_{\mathrm s}=1$, $K=0.2$, $a=1$, $\gamma=0.1$.
}
 \label{fig:1}
\end{figure}

\begin{figure}
 \centering
 \includegraphics[width=80mm]{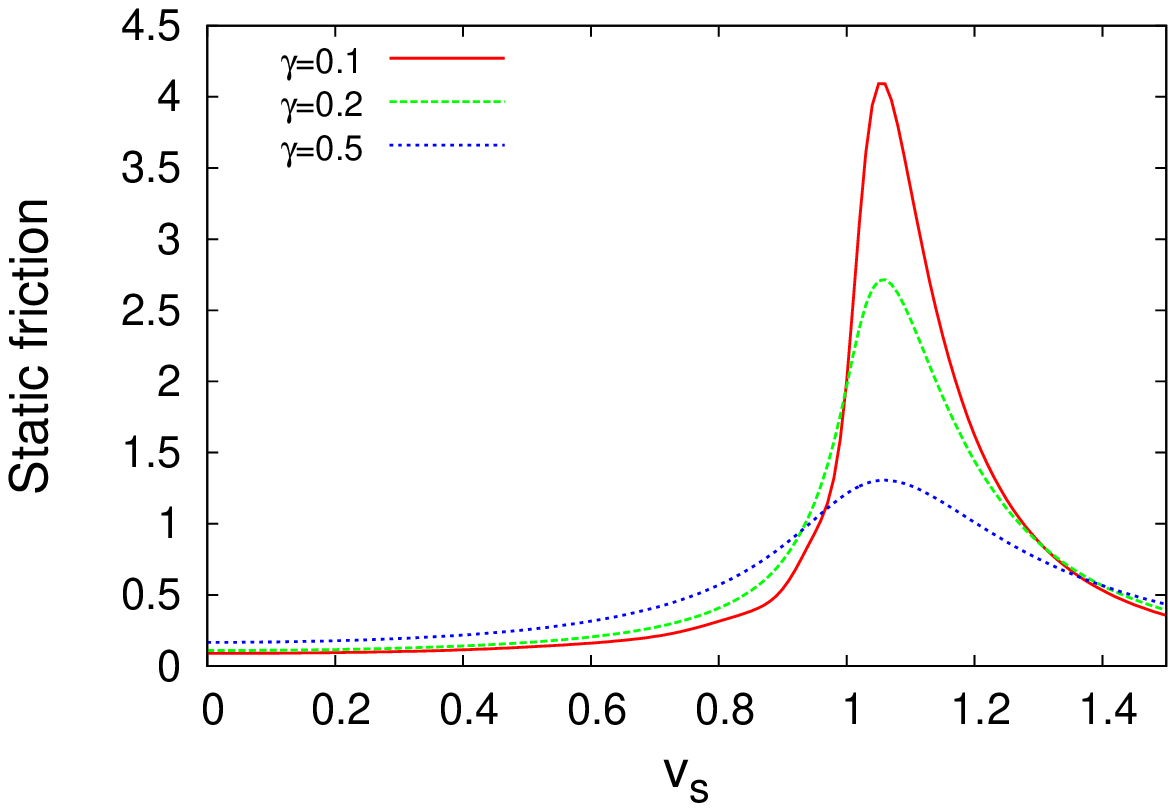}
 \caption{
Static friction $\tilde{\Gamma}^{(0)}(\omega=0)$ for the weak-friction limit as a function of mean velocity for $\gamma$ = 0.1 (solid line), 0.2 (dashed line), and 0.5 (dotted line).
 Parameters: $\lambda=1$, $c_{\mathrm s}=1$, $K=0.2$, $a=1$.
}
 \label{fig:2}
\end{figure}

\subsection{Correlation function of velocity fluctuation}

\subsubsection{The weak-friction limit}

Let us consider the weak-friction limit where the damping constant of the elastic medium $\gamma$ is considerably smaller than the frequency of the mode that most strongly interacts with the particle.
Since the mode whose wavelength is nearly equal to the interaction range $a$ is most important, this condition is written as
\begin{equation}
\gamma \ll \omega_{k=1/a}=\sqrt{K+(c_s/a)^2}.
\end{equation}
Then, $G_k(t)$ is given by
\begin{equation}
G_k(t)=\lambda^2 \frac{k^4}{\omega_k^2} \: \tilde{V}_{\rm int}(k)^2 \cos \omega_kt  e^{-\frac{\gamma}{2}t},
\end{equation}
because $\alpha_{k\pm} \simeq \pm i\omega_k-\gamma/2$ for the mode with wavenumber $k \sim 1/a$. 
$\Gamma_{\rm eff}(t)$ is calculated by using Eqs.~(25), (28) and (30).

In what follows, we take the Gaussian function as the interaction potential
\begin{equation}
V_{\rm int}(x)=\exp \left[-\frac{x^2}{2a^2}\right].
\end{equation}
We display the friction kernel $\Gamma_{\rm eff}(t)$ for different values of the mean velocity $v_{\rm s}$ in Fig.~1. 
We observe that the closer to the sound velocity $c_{\rm s}=1.0$ the mean velocity is, the longer the relaxation time of the friction kernel becomes. 
We also show the static friction $\tilde{\Gamma}^{(0)}(\omega=0)$ as a function of $v_{\rm s}$ for different values of $\gamma$ in Fig.~2. 
The static friction increases with the mean velocity below the sound velocity.
Since the correlation of the random force $R(t)$ is given by $\Gamma^{(0)}(t)$, this means that $R(t)$ is more strongly correlated in NESS than in the equilibrium state. 
In Fig.~3, $C_{\rm ss}(\omega)$ is shown for different values of the mean velocity. In NESS, a pronounced peak is observed.
This peak is regarded as the consequence of increase in the relaxation time of the friction kernel.

\begin{figure}
 \centering
 \includegraphics[width=80mm]{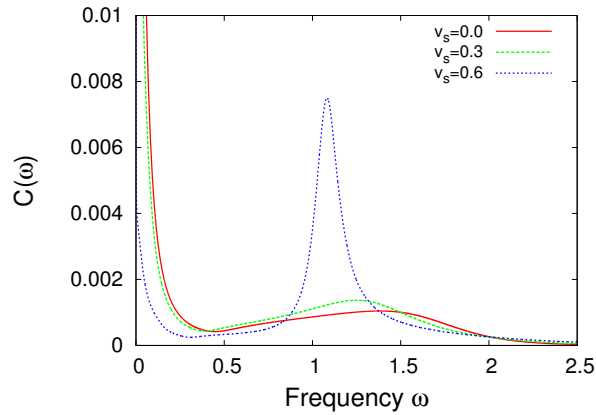}
 \caption{
Power spectrum of velocity fluctuation $C(\omega)$ of the weak-friction limit for the mean velocity $v_{\rm s}$ = 0.0 (solid line), 0.3 (dashed line), and 0.6 (dotted line). 
 Parameters: $\lambda=1$, $c_{\rm s}=1$, $a=1$, $M=1$, $\gamma=0.1$, $K=0.2$, $T=0.001$. 
}
 \label{fig:3}
\end{figure}

\begin{figure}
 \centering
 \includegraphics[width=80mm]{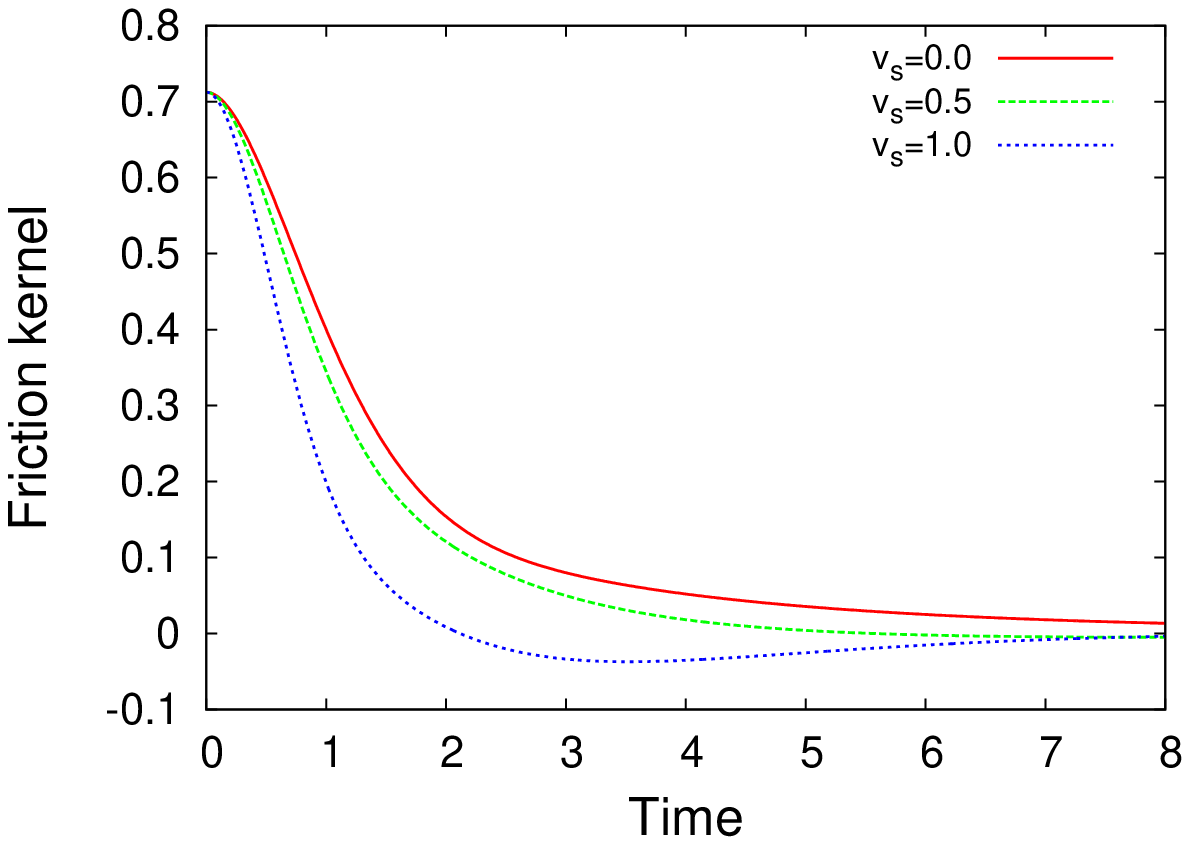}
 \caption{
Friction kernel $\Gamma_{\rm eff} (t) $ for the strong-friction limit for the mean velocity $v_{\rm s}$ = 0.0 (solid line), 0.5 (dashed line), and 1.0 (dotted line). 
Parameters: $\lambda=1$, $g=1$, $K=0.2$, $a=1$, $\gamma=2.0$.
}
 \label{fig:4}
\end{figure}

\subsubsection{The strong-friction limit}

We next consider the strong-friction limit
\begin{equation}
\gamma \gg \omega_{k=1/a}=\sqrt{K+(c_s/a)^2}.
\end{equation}
Noting that $\alpha_{k\pm} \simeq -\omega_k^2/\gamma,\:-\gamma$, $G_k(t)$ becomes
\begin{equation}
G_k(t)=\lambda^2 \frac{k^4}{\omega_k^2} \: \tilde{V}_{\rm int}(k)^2  e^{-\frac{\omega^2_k}{\gamma}t}.
\end{equation}
Let us investigate the asymptotic behavior of $\Gamma_{\rm eff}(t)$ in the large-$t$ region.
Since the integrand of Eq.~(25) has dominant contribution in $k \sim 0$ in the large-$t$ region, one can replace $\tilde{V}_{\rm int}(k)$ with a constant value $\tilde{V}_{\rm int}(0)$. 
Moreover, $\Gamma^{(1)}(t)$ has the contribution of $O((v_{\rm s}/c_{\rm s})^2)$ for a small velocity $v_{\rm s} \ll a \gamma, c_{\rm s} $.
Consequently, we describe the large-$t$ behavior of $\Gamma_{\rm eff}(t)$ as the following:
\begin{equation}
\Gamma_{\rm eff}(t) \sim \frac{1}{K} P(t) \exp \left[-\left(\frac{K}{\gamma}+\frac{\gamma v_{\mathrm s}^2}{4g} \right)t  \right]+O \left((v_{\rm s}/c_{\rm s})^2 \right),
\end{equation}
where $P(t)$ denotes a polynomial with fractional power of $t$. This result is independent of the detailed form of the interaction potential $V_{\rm int}(x)$.
The relaxation time of the friction kernel decreases with increasing mean velocity.

In the gapless dispersion limit $K \ll g/a^2$, we have the asymptotic form of $\Gamma_{\rm eff}(t)$:
\begin{equation}
\nonumber \Gamma_{\rm eff}(t) \sim t^{-3/2},\:\:\:\:\:{\rm for}\:\: v_{\rm s}=0,
\end{equation}
and
\begin{equation}
\nonumber \Gamma_{\rm eff}(t) \sim Q(t)\exp \left[-\frac{\gamma v_{\mathrm s}^2t}{4g}  \right],  \:\:\:\:\:{\rm for} \:\: v_{\rm s} \neq 0,
\end{equation}
in the large-$t$ region. Thus, $\Gamma_{\rm eff}(t) $ exhibits a power law decay in equilibrium but an exponential decay in NESS if $K \ll g/a^2$.

\begin{figure}
 \centering
 \includegraphics[width=80mm]{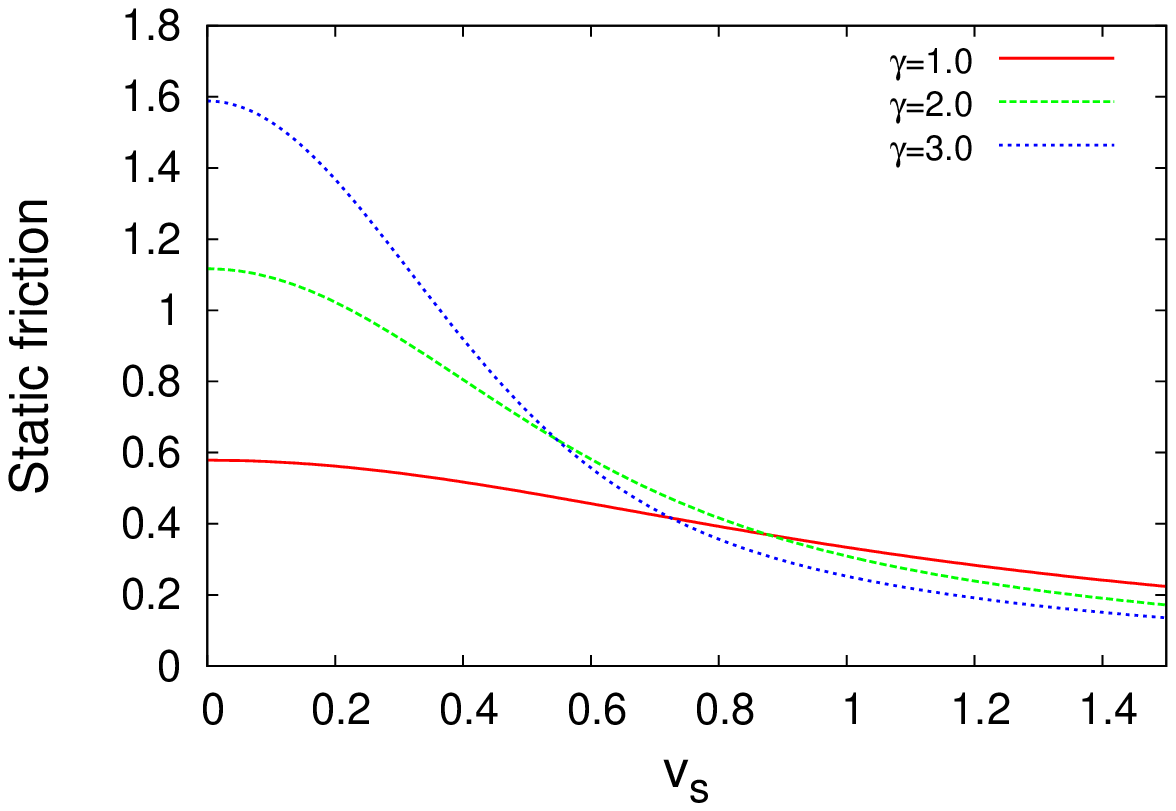}
 \caption{
Static friction $\tilde{\Gamma}^{(0)}(\omega=0)$ for the strong-friction limit as a function of mean velocity for $\gamma$ = 1.0 (solid line), 2.0 (dashed line), and 3.0 (dotted line).
 Parameters: $\lambda=1$, $g=1$, $K=0.2$, $a=1$.
}
 \label{fig:5}
\end{figure}

\begin{figure}
 \centering
 \includegraphics[width=80mm]{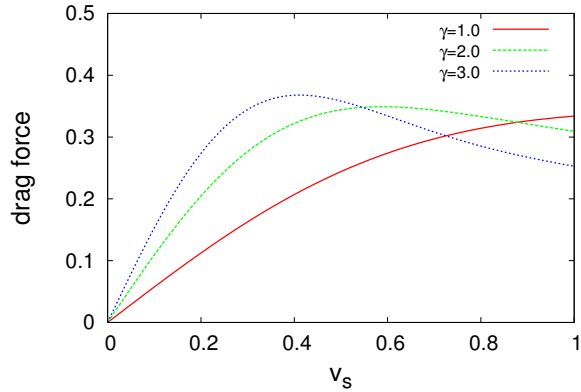}
 \caption{
Drag force $\tilde{\Gamma}^{(0)}(\omega=0) v_s$ for the strong-friction limit as a function of mean velocity for $\gamma$ = 1.0 (solid line), 2.0 (dashed line), and 3.0 (dotted line).
 Parameters: $\lambda=1$, $g=1$, $K=0.2$, $a=1$.
}
 \label{fig:6}
\end{figure}

We display the friction kernel $\Gamma_{\rm eff}(t)$ for different values of the mean velocity $v_{\rm s}$ in Fig.~4. 
We observe that in NESS the relaxation time becomes shorter than that in equilibrium as predicted by Eq.~(45).
This tendency is opposite to that of the weak-friction limit. 
We also show the static friction $\tilde{\Gamma}^{(0)}(\omega=0)$ as a function of $v_{\rm s}$ in Fig.~5. 
The result indicates that $\tilde{\Gamma}^{(0)}(\omega=0)$ decreases monotonically with increasing  $v_{\rm s}$. 
The drag force that is given by $\tilde{\Gamma}^{(0)}(\omega=0) v_{\rm s}$ is shown in Fig.~6 as a function of $v_{\rm s}$.
For large values of $\gamma$, there is a region in which the drag force decreases with increasing mean velocity, and thus, no NESS exists in such a region.
In Fig.~7, $C_{\rm ss}(\omega)$ is shown for different values of the mean velocity. 
We observe a disappearance of a peak with increasing mean velocity. This disappearance is regarded as the consequence of decrease in the relaxation time of the friction kernel.

\subsection{Kinetic temperature}

Let us calculate the kinetic temperature from Eq.~(39).
We show $T_{\rm K}$ as a function of mean velocity in Fig.~8.
For the strong-friction limit ($\gamma=2.0$), the kinetic temperature increases with the mean velocity.
On the other hand, for the weak-friction limit ($\gamma=0.1$), the kinetic temperature decreases in the small-$v_{\rm s}$ region. 
Furthermore, it increases in the large-$v_{\rm s}$ region.

The explanation of these results is as follows. 
For the strong-friction limit, the static friction is a decreasing function of mean velocity (Fig.~5).
Thus, the fluctuation in velocity is enhanced, and subsequently, the kinetic temperature increases.
For the weak-friction limit, the static friction increases with the mean velocity below the sound velocity (Fig.~2).
Thus, the fluctuation in velocity is suppressed, and subsequently, the kinetic temperature decreases. 
Note that this result does not violate the second law of thermodynamics.
The nonmonotonic behavior may be due to the inertial effect related to the second term of the denominator of the integrand in Eq.~(39).

\begin{figure}
 \centering
 \includegraphics[width=80mm]{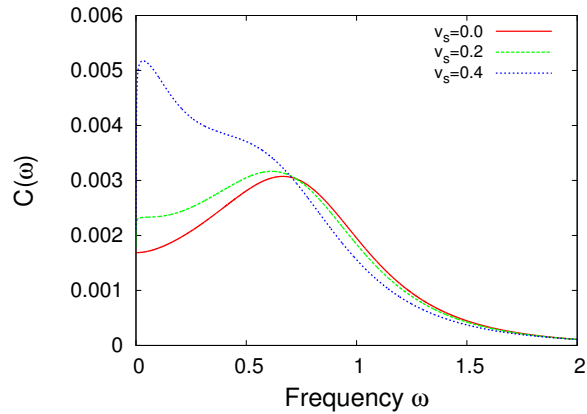}
 \caption{
Power spectrum of the velocity fluctuation $C(\omega)$ of the strong-friction limit for the mean velocity $v_{\rm s}$ = 0.0 (solid line), 0.2 (dashed line), and 0.4 (dotted line). 
 Parameters: $\lambda=1$, $g=1$, $a=1$, $M=1$, $\gamma=2$, $K=0.2$, $T=0.001$. 
}
 \label{fig:7}
\end{figure}

\section{Effective temperature for NESS}

In this section, we calculate the effective temperature, which is defined as the ratio of the correlation function of the velocity fluctuation to the linear response function.
We discuss the relationship between the effective temperature and the kinetic temperature.
As the first step, we define the linear response function in NESS.
Suppose the particle exists in NESS with a mean velocity $v_{\rm s}$. We apply a time dependent weak probe force $f(t)$ in addition to the driving force $F$ for 
$t \geq t_0$. Then, the linear response function $R_{\rm ss}(t)$ in NESS is defined as
\begin{equation}
\Delta v(t)=\left< v \right>_{F+f}-\left< v \right>_F=\int_{t_0}^t R_{\rm ss}(t-s)f(s){\rm d}s
\end{equation}
where $\left< ... \right>$ denotes the ensemble average with respect to realization of the noise $\xi(t)$. 
We define the frequency-dependent effective temperature $T_{\rm eff}(\omega)$ as
\begin{equation}
k_{\rm B}T_{\rm eff}(\omega)=\frac{C_{\rm ss}(\omega)}{2 {\rm Re}\tilde{R}_{\rm ss}(\omega)}.
\end{equation}
where $\tilde{R}_{\rm ss}(\omega)$ denotes the Fourier-Laplace transform of $R_{\rm ss}(t)$.
From the effective Langevin equation (29), ${\rm Re}\tilde{R}_{\rm ss}(\omega)$ is given by
\begin{equation}
{\rm Re}\tilde{R}_{\rm ss}(\omega)=\frac{{\rm Re}\tilde{\Gamma}_{\rm eff}(\omega)}{{\rm Re}\tilde{\Gamma}_{\rm eff}(\omega)^2+(M\omega+{\rm Im}\tilde{\Gamma}_{\rm eff}(\omega))^2}.
\end{equation}
Therefore, we have
\begin{equation}
\frac{T_{\rm eff}(\omega)}{T}=\frac{\:{\rm Re}\tilde{\Gamma}^{(0)}(\omega)}{\:{\rm Re}\tilde{\Gamma}_{\rm eff}(\omega)}.
\end{equation}

\begin{figure}
 \centering
 \includegraphics[width=80mm]{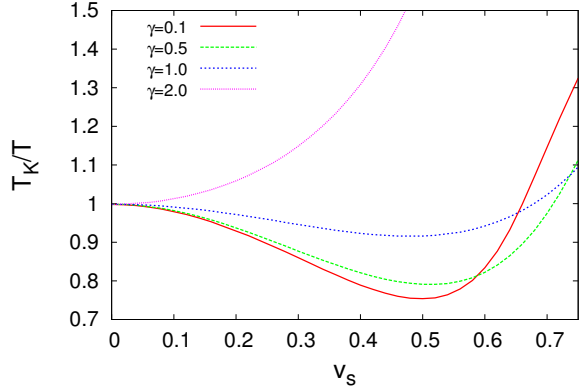}
 \caption{
Kinetic temperature $T_{\rm K}$ as a function of mean velocity for different values of $\gamma$. 
Parameters: $\lambda=1$, $g=1$, $a=1$, $M=1$, $\gamma=0.1,\:0.5,\:1.0,\:2.0 $ (from bottom to top at $v_{\rm s}=0.5$), $K=0.2$, $T=0.001$.
}
 \label{fig:8}
\end{figure}

We consider the case wherein the time scale of the dynamics of the elastic medium is considerably faster than that of the particle.
The time scale of the dynamics of the elastic medium and that of the particle are given by
\begin{equation}
\tau_{\rm E}=\gamma^{-1},
\end{equation}
and
\begin{equation}
\tau_{\rm P}=\frac{M}{\int^{\infty}_0 \Gamma_{\rm eff}(t) {\rm d}t},
\end{equation}
respectively. Thus, the above condition is written as
\begin{equation}
\tau_{\rm E} \ll \tau_{\rm P}.
\end{equation}
The expression (39) is rewritten as
\begin{equation}
k_{\rm B}T_{\rm K}=\int \frac{{\rm d}\omega}{2\pi} \frac{2 M k_{\rm B}T_{\rm eff}(\omega) {\rm Re}\tilde{\Gamma}_{\rm eff}(\omega)}{{\rm Re}\tilde{\Gamma}_{\rm eff}(\omega)^2+(M\omega+{\rm Im}\tilde{\Gamma}_{\rm eff}(\omega))^2}.
\end{equation}
The numerator and denominator of the integrand vary over the scale of $\tau_{\rm E}^{-1}$ and $\tau_{\rm P}^{-1}$, respectively.
Thus, $T_{\rm eff}(\omega)$ can be put out of the integral with $\omega=0$ because of the condition (52).
Noting that 
\begin{equation}
\int \frac{{\rm d}\omega}{2\pi} \frac{2 M \:{\rm Re}\tilde{\Gamma}_{\rm eff}(\omega)}{{\rm Re}\tilde{\Gamma}_{\rm eff}(\omega)^2+(M\omega+{\rm Im}\tilde{\Gamma}_{\rm eff}(\omega))^2}=1,
\end{equation}
we obtain the following formula:
\begin{equation}
\frac{T_{\rm K}}{T}=\frac{T_{\rm eff}(\omega=0)}{T}=\frac{\tilde{\Gamma}^{(0)}(\omega=0)}{\tilde{\Gamma}_{\rm eff}(\omega=0)}.
\end{equation}
as our central result.
The effective static friction $\tilde{\Gamma}_{\rm eff}(\omega=0)$ is given by
\begin{equation}
\tilde{\Gamma}_{\rm eff}(\omega=0)=\tilde{\Gamma}^{(0)}(\omega=0)+v_{\rm s}\frac{\partial}{\partial v_{\rm s}}\tilde{\Gamma}^{(0)}(\omega=0).
\end{equation}
Therefore, the effective temperature associated with the time scale of the particle is identical to the kinetic temperature.
This result is consistent with those reported in Refs.~[18] and [19], wherein the effective temperature is calculated for the Langevin model with a tilted periodic potential.
With respect to Eq.~(55), it is remarkable that the kinetic temperature does not depend on the mass of the particle.
In other words, the kinetic temperature is not equal to the bath temperature even in the limit of $\tau_{\rm P}/\tau_{\rm E} \to \infty $, 
for which the environment is expected to behave as an ideal heat bath. 
It is also important that this formula is written by measurable quantities. 
The numerator of the right-hand side is the static friction, which is simply the ratio of the external force to the mean velocity. 
The denominator of the right-hand side represents the effective static friction for the fluctuation around NESS. It can be determined by measuring the linear response in NESS.
These quantities do not coincide in the presence of nonlinear friction. 
This expression indicates that $T_{\rm K}<T$ when $v''_{\rm s}(F)<0$ and $T_{\rm K}>T$ when $v''_{\rm s}(F)>0$.
Thus, the kinetic temperature is determined by the curvature of the force-velocity curve.
These results are consistent with those observed in Fig.~8.
Note that this formula is correct when the condition (52) is satisfied. The nonmonotonic behavior shown in Fig.~8 for the weak-friction limit is not derived from this formula.

We next consider the opposite case where
\begin{equation}
\tau_{\rm E} \gg \tau_{\rm P}.
\end{equation}
In Eq.~(53), $T_{\rm eff}(\omega)$ can be put out of the integral with $\omega \tau_{\rm E} \to \infty$ because of the condition (57).
Thus, we have
\begin{equation}
\frac{T_{\rm K}}{T}=\lim_{\omega \tau_{\rm E} \to \infty} \frac{T_{\rm eff}(\omega)}{T}=1.
\end{equation}
We note that the heat bath temperature and the kinetic temperature appear as the limiting cases of the frequency-dependent effective temperature.

\section{Summary}

We investigated a simple model that describes the dynamics of a particle interacting with a spatially extended environment.
The effective Langevin equation for nonequilibrium steady states driven by a constant external force is derived by eliminating the degrees of freedom of the environment.
As a result, we obtained the following observations.

The model exhibits different behaviors in two limiting cases: the weak-friction limit $\gamma \ll \omega_{k=1/a}$ and the strong-friction limit $\gamma \gg \omega_{k=1/a}$.
In the weak-friction limit, the relaxation time of the friction kernel $\Gamma_{\rm eff}(t)$ increases with the mean velocity of the particle. Conversely, in the strong-friction limit,
we found that the relaxation time decreases with increasing mean velocity. This behavior leads to the negative slope observed for the drag force-velocity curve.

The kinetic temperature $T_{\rm K}$ of the particle in NESS is calculated as a function of mean velocity. In the strong-friction limit, the kinetic temperature monotonically increases with the mean velocity. 
On the other hand, in the weak-friction limit, the kinetic temperature decreases and exhibits nonmonotonic dependence. From the simple expressions (55) and (56), we note that 
the decrease in $T_{\rm K}$ stems from the increase in the static friction $\tilde{\Gamma}^{(0)}(\omega=0)$.

We calculated the effective temperature, which is defined as the ratio of the correlation function of the velocity fluctuation to the linear response function.
It is shown that the effective temperature associated with the time scale of the particle is identical to the kinetic temperature 
if the time scale of the environment $\tau_{\rm E}$ and that of the particle $\tau_{\rm P}$ are well separated. 
This result indicates that the fluctuation-dissipation relation recovers in NESS by replacing the heat bath temperature $T$ 
with the kinetic temperature $T_{\rm K}$ for the time scale of the particle.
It is consistent with observations for driven glassy systems [12-15].
A similar behavior is also observed in athermal systems.
For a Brownian particle in driven granular matter, it is shown experimentally that the effective temperature is identical to the granular temperature in the low-frequency region [20].
Furthermore, we derived a simple formula Eq.~(55) that relates the kinetic temperature with the curvature of the driving force-mean velocity curve.
This relation can be investigated by experiment or detailed molecular dynamics simulation for more realistic systems.
At present, the universality of this result is not clear. Thus, further research is required to elucidate it.

\begin{acknowledgements}
We express special thanks to Shin-ichi Sasa for continuous discussions.
We also thank T. Kawakatsu for useful discussions.
The present study was supported by KAKENHI Nos. 22340109 and 25103002 and by the JSPS Core-to-Core program ``Non-equilibrium dynamics of soft-matter and information.''
\end{acknowledgements}

\end{document}